\documentstyle[epsf]{disk}
\begin{document}

\title{An unstable inner disk in Cyg X-1 ?}

\author{Chris DONE and Piotr T.\ \.{Z}YCKI\\
{\it Department of Physics, University of Durham,
South Road, Durham DH1 3LE, England, U.K.}} 

\maketitle

\section*{Abstract}

We present the first significant detection of relativistic smearing of the
X--ray reflection spectrum from the putative accretion disk in the low/hard
state of Cyg X--1.  The ionization state, and amount of relativistic smearing
are simultaneously constrained by the X--ray spectra, and we conclude that the
disk is not strongly ionised, does not generally extend down to the last stable
orbit at 3 Schwarzschild radii and covers rather less than half the sky as seen
from the X--ray source. These results are consistent with a geometry where the
optically thick disk truncates at a few tens of Schwarzschild radii, with the
inner region occupied by the X--ray hot, optically thin(ish) plasma. Such a
geometry is also inferred from previous studies of the reflected spectrum in
Galactic Black Hole transient sources, and from detailed considerations of the
overall continuum spectral shape, suggesting that there is some robust
instability which disrupts the inner accretion disk in low/hard state Galactic
Black Holes. 

\section{Introduction}

Some of the strongest evidence for the existence of black holes has come from
recent ASCA (0.6--10 keV) X--ray observations of the shape of the iron K$\alpha$
line in Active Galactic Nuclei (AGN). AGN typically produce copious hard X--ray
emission, and the iron line is formed from fluorescence as these X--rays
illuminate the infalling material. The combination of Doppler effects from the
high orbital velocities and strong gravity close to the black hole gives
the line a characteristically skewed, broad profile (Fabian et al., 1989). This
has been unambiguously identified in ASCA data from the AGN
MCG--6--30--15, where the line width implies that the accretion
disk extends down to {\it at least} $6R_g$ ($R_g=GM/c^2$), the last stable
orbit in a Schwarzschild metric (Tanaka et al 1995).

As well as producing the line, some fraction of the illuminating hard X--rays
are reflected from the accretion flow, producing a characteristic continuum
spectrum.  The amplitude of the line and reflected continuum depend on the
amount of material being illuminated by the hard X--rays, its inclination,
elemental abundances and ionization state (e.g.\ Lightman \& White 1988; George
\& Fabian 1991). The amount of reflection and line seen in AGN are
consistent with a power law X--ray spectrum illuminating an optically thick,
(nearly) neutral disk, which subtends a solid angle of $\sim 2\pi$ (e.g. Pounds
et al 1990).

This contrasts with the situation in the Galactic Black Hole Candidates
(GBHC). These are also thought to be powered by accretion via a disk onto a
black hole, and, in their low/hard state, show spectra which are rather
similar to those from AGN. However, the amount of reflection and iron line is
much less than would be expected from an accretion disk which subtends a solid
angle of $2\pi$ (e.g. Gierli\'{n}ski et al 1997),
and the detected line is narrow, with no obvious broad component
(e.g. Ebisawa et al 1996, hereafter E96). There are (at least)
two possible explanations for this: firstly that it is an artifact of a
difference in ionization state of the disk between GBHC and AGN, or secondly
that there is a real difference in geometry between the supermassive 
and stellar mass objects.

Ionization differences seems at first sight to be an extremely attractive
option. For accretion at the same fraction of Eddington, the disk temperature
should scale as $M^{-1/4}$. The GBHC inner disk is then expected to be a factor
of $\sim 30$ hotter than in AGN, and the higher expected ionization state from
the thermal ion populations gives a reflected continuum and associated iron line
that can be very different to that from a neutral disk (e.g.  Ross, Fabian \&
Brandt 1996).  Fits with photo--ionised reflected continuua show that the
ionization state of the reflector is generally rather low (Done et al 1992;
Gierli\'{n}ski et al 1997), arguing against such models. However, relativistic
shifts could affect the derived ionization states (e.g. Ross et al., 1996). 
Here we fit relativistically smeared, ionized models of the
reflected continuum to the low/hard state spectra from Cyg X--1 and compare this
to data from other GBHC and AGN.

\section{Spectral Fitting}

We use a reflection model described in detail in $\dot{\rm Z}$ycki et al
(1998b), where the iron K$\alpha$ line strength is calculated self consistently
with the properties of the reflected continuum, and relativistic effects are
then applied to this total reprocessed spectrum. This gives a much more powerful
approach than fitting the (relativistically smeared) line and (unsmeared)
reflected continuum separately.  The free parameters are the solid angle
subtended by the reflector to the X--ray source, $\Omega/2\pi$, (an isotropic
X--ray source above a flat disk has $\Omega/2\pi=1$), ionisation state, $\xi$,
and inner radius of the disk $R_{in}$. We use Morrison \& McCammon (1983)
abundances, with only iron free to vary from solar Fe/H($=3.3\times 10^{-5}$).

\subsection{EXOSAT GSPC}

The EXOSAT GSPC data from Cyg X--1 give some of the best spectra to date from
this object, with the broad 2--20 keV energy range of GINGA data but resolution
comparable to that of the ASCA GIS. We use the 5 GSPC spectra of Done et al.,
(1992). The residuals to a simple power law are shown in Figure 1a, together
with the residuals to a simple power law fit to the GINGA--12 AGN spectrum of
Pounds et al (1990).  Contrary to claims by Ross et al (1996), Cyg X--1 is not
dominated by the edge structure, but also shows a strong excess at the iron line
energy. {\it Both} line and edge appear equally diminished in Cyg X--1 compared
to AGN.

\begin{figure}[!h]
\begin{center} \leavevmode
\hbox{%
\epsfysize=0.45\textwidth
\epsffile[28 34 517 501]{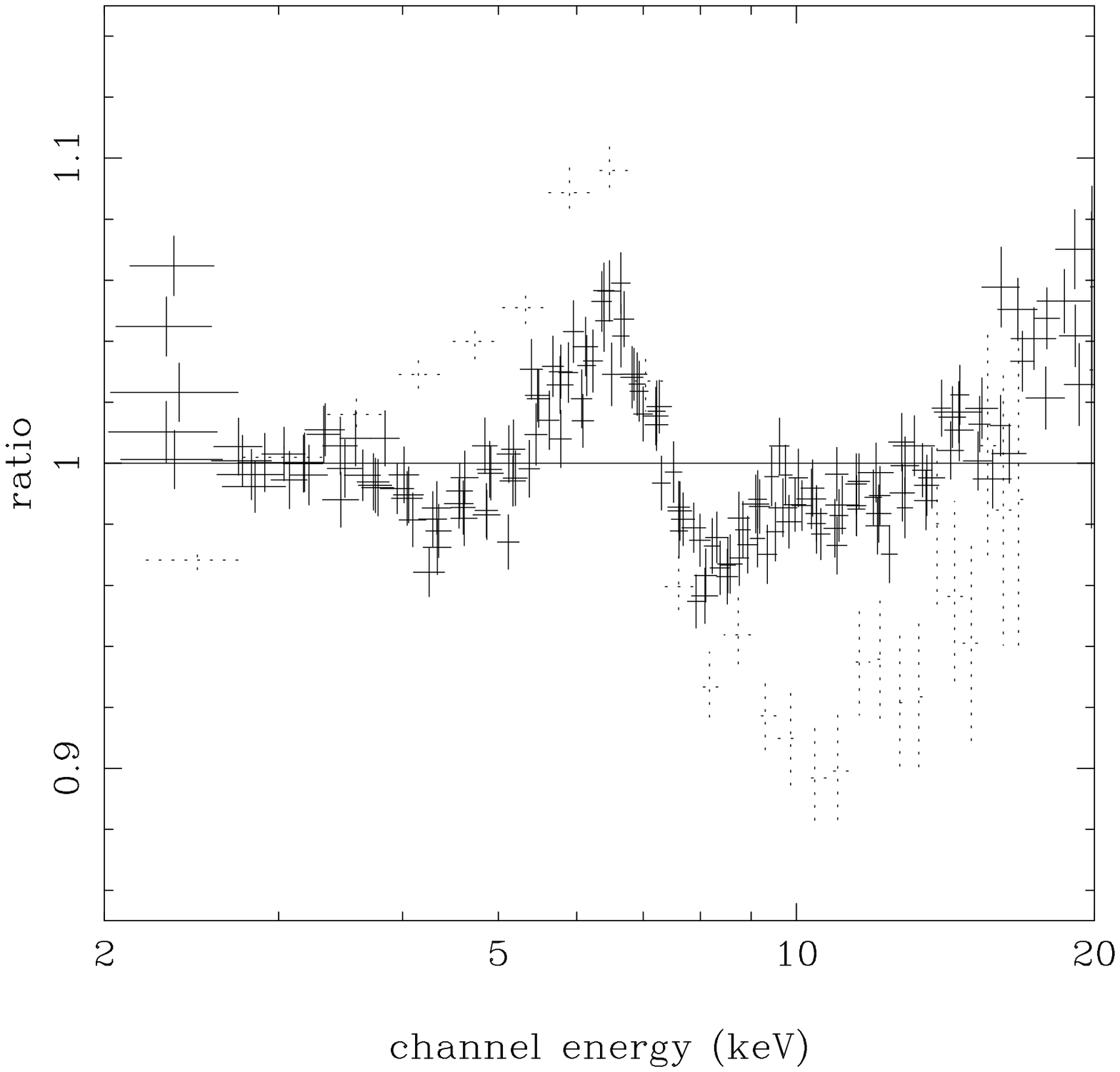}
\centering \leavevmode
\epsfysize=0.45\textwidth
\epsffile[28 34 517 501]{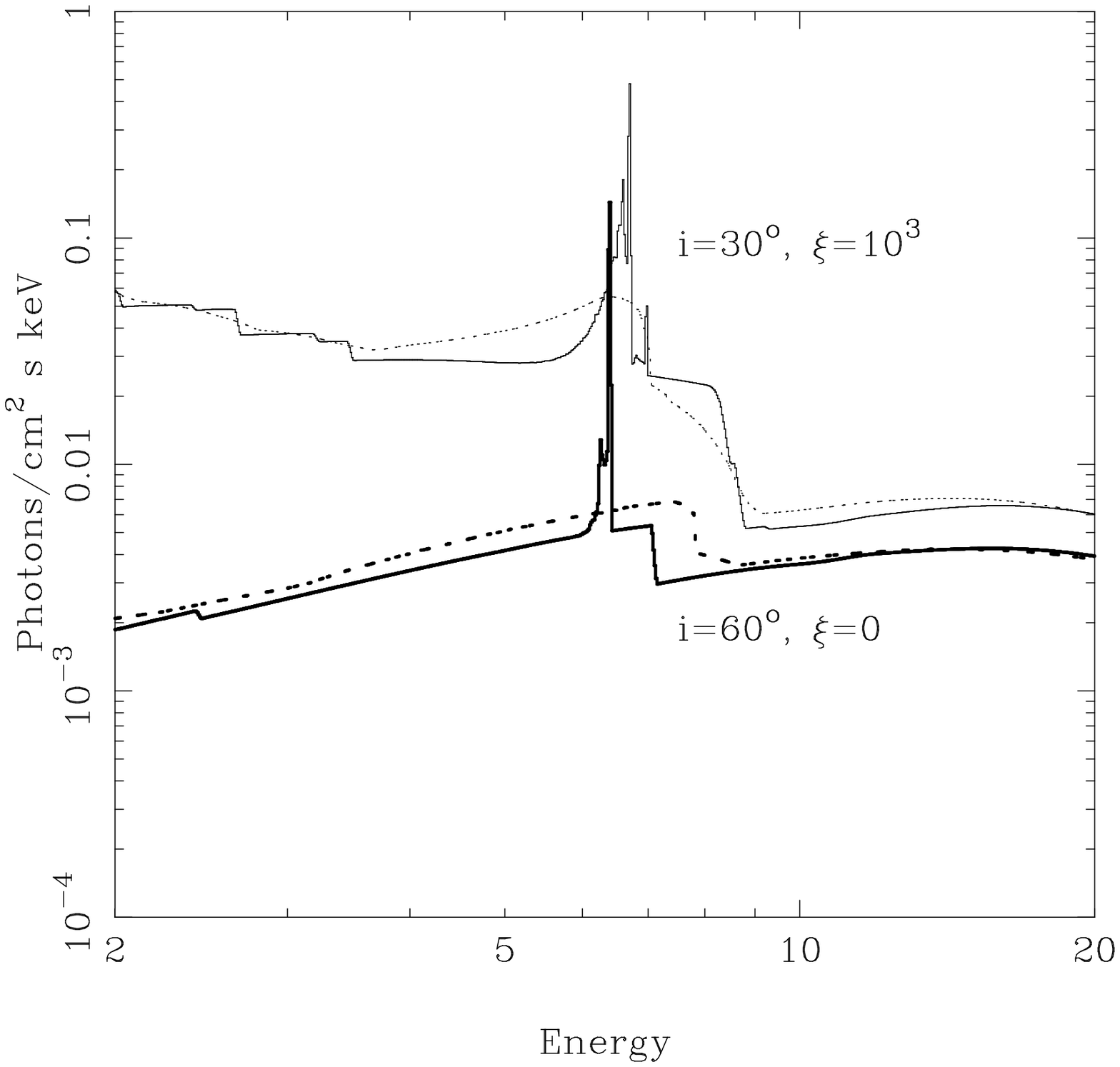}}
\end{center}
\caption{
The left hand panel shows the ratio of the EXOSAT GSPC spectra of Cyg X--1 to
a power law model, while the dotted line shows this for the 
GINGA--12 AGN spectrum (Pounds et al 1990). 
The right hand panel shows how relativistic smearing and ionization
can distort the observed
reprocessed spectrum. The thick solid line is the reflected (line and continuum)
resulting from power law illumination of a neutral slab inclined at $60^\circ$,
while the dotted line shows the distortions induced by relativistic effects if
the reprocessor is a disk which extends down to the $6R_g$. The thin solid and
dotted lines show results for highly ionised ($\xi=10^3$) material inclined at
$30^\circ$.
}
\end{figure}

We fit only the data above 4 keV to avoid problems with the soft X--ray excess,
and fix the galactic column at $6\times 10^{21}$ cm$^{-2}$.
The 5 GSPC spectra are fit simultaneously in XSPEC, although only the reflection
parameters of inclination and iron abundance are constrained to be equal across
all the datasets. A power law and lightly ionised reprocessed spectrum inclined
at $30^\circ$ gives a good fit to the data ($\chi^2_\nu=611/615$), with a
derived iron abundance which is close to solar. However, a significantly better
fit is obtained if the reprocessed spectrum is relativistically smeared, giving
$\chi^2_\nu=535/604$. The derived inner radius of the disk in each spectrum is
generally {\it inconsistent} with the disk extending down to the last stable
orbit, assuming an illumination which is $\propto r^{-3}$, and the derived
ionization is rather low with typical values of $\xi\sim 30$, corresponding to
FeXII. The iron abundance is now derived to be twice solar, consistent with the
abundance of iron inferred for the stellar wind of the companion star from the
strong edge feature seen in the absorbed `dip' spectra (Kitamoto et al., 1984).
Relativistic smearing allows more iron line to be present, since it is broad
rather than narrow, so it is much less observable (see figure 1b). These fits
also include an additional (neutral and unsmeared) reprocessed spectrum to allow
for a contribution from the companion star and/or outer disk (Basko 1978); such
narrow features are detected in high resolution data (e.g. E96) although they
are not significant in the EXOSAT data.

\subsection {ASCA GIS and SIS}

This model (power law, relativistically smeared reprocessed spectrum, and
unsmeared reprocessed spectrum) was also fit to the 4--10 keV ASCA GIS and SIS
datasets of E96. The 8 ASCA GIS spectra can again be fit simultaneously, with
only the refected parameters of abundance and inclination tied across the
datasets. A power law and an unsmeared reflected spectrum inclined at $30^\circ$
give $\chi^2_\nu=951.9/647$. The derived reflected spectra are all lightly
ionised, again with typical $\xi\sim 30$. The situation changes dramatically
when relativistic smearing is included, along with an unsmeared, unionised reflected
component from the companion star. The fit is clearly statistically
significantly better -- $\chi^2_\nu$ drops to $753/632$. However, the derived
ionisation state of the relativistic reflector is very high, typically with
$\xi\sim 10^{3-4}$. Figure 2a shows the best fit to one of the ASCA GIS spectra
(3). The ionisation is so extreme that the 'edge' feature seen in the data is
actually modelled by the end of the
relativistically smeared ionised line, and the real edge is shifted out of the
observed energy range. Similar results are seen for most of the
ASCA GIS and SIS spectra.

\begin{figure}[!h]
\begin{center} \leavevmode
\hbox{%
\epsfysize=0.45\textwidth
\epsffile[28 34 517 501]{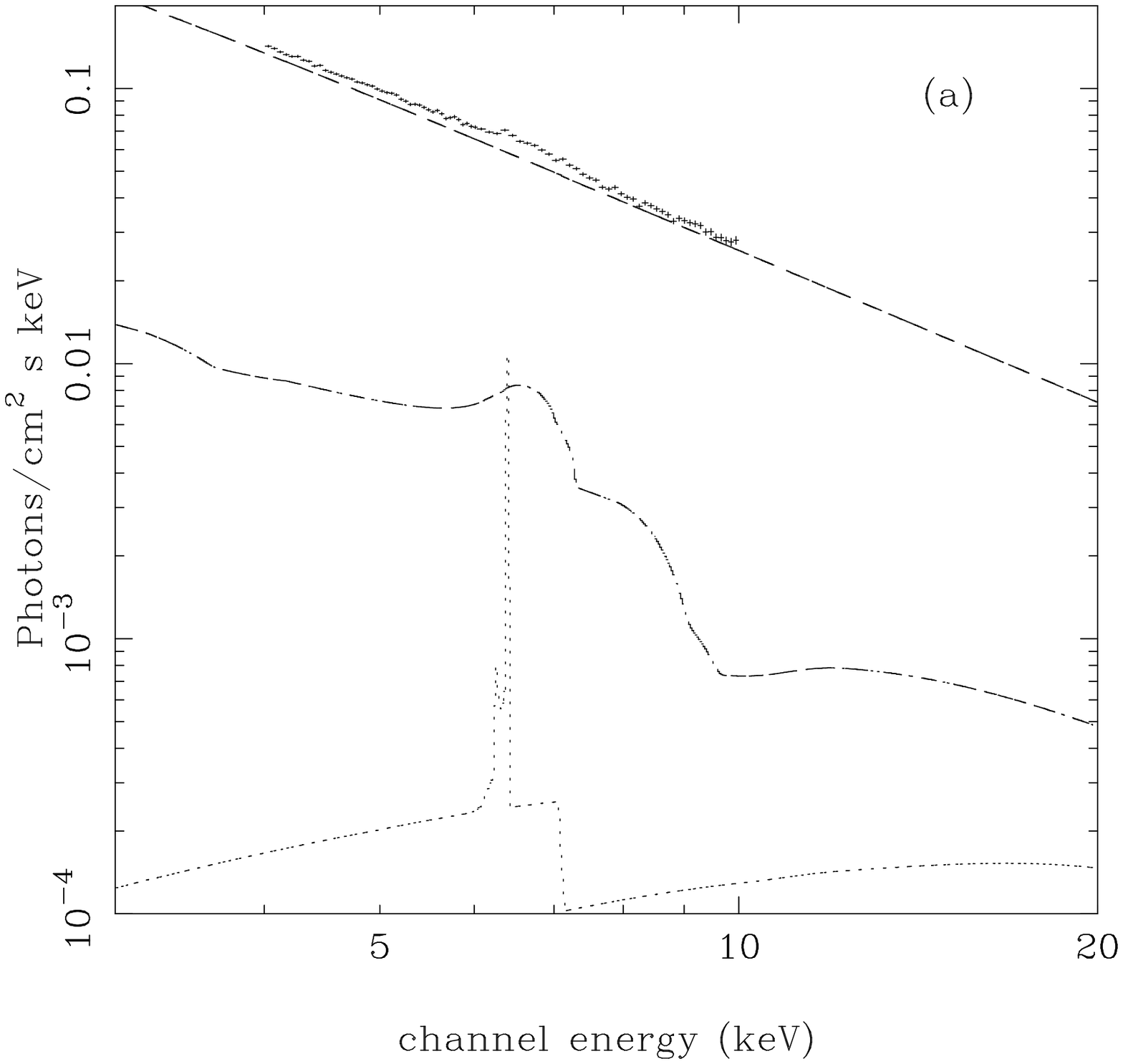}
\centering \leavevmode
\epsfysize=0.45\textwidth
\epsffile[28 34 517 501]{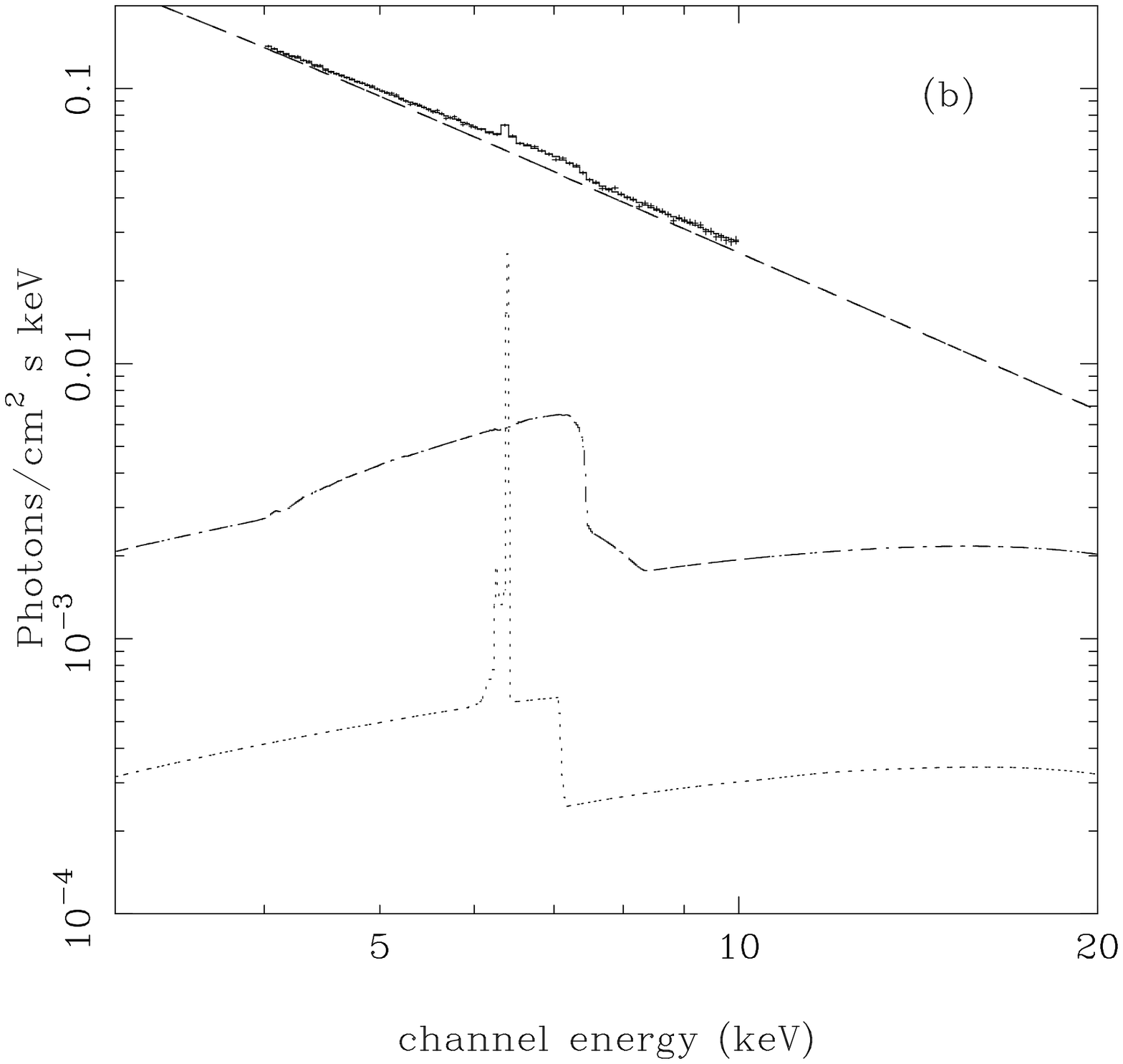}}
\end{center}
\caption{
Panel (a) shows the best fit to the ASCA GIS 3 spectrum with a
relativistic disk inclined at $30^\circ$. The derived ionisation state of the
reflector is extreme with $\xi= 3.0\pm 2.4\times 10^4$, 
for solid angle $\Omega/2\pi=0.06^{+0.03}_{-0.02}$ and inner disk
radius of $58_{-31}^{+110}R_g$ ($\chi^2_\nu=92/79$). The right hand panel shows
the significantly better ($\chi^2_\nu=69.8/79$) higher inclination $53^\circ$
solution for the GIS 3 data, which has $\Omega/2\pi=0.53^{+0.09}_{-0.17}$, 
an inner radius of $9.9_{-1.9}^{+6}R_g$ and $\xi=0^{+17}$. This corresponds 
much better to the EXOSAT solutions, which always pick a low ionisation
reflector, irrespective of inclination.
}
\end{figure}

\subsection{Inclination}

This general mismatch between the ionization of the reprocessor derived from
ASCA and EXOSAT data is suggestive of a systematic problem in the spectral model
fitting, most likely due to the difference in bandpass. In EXOSAT the high
energy continuum shape of the reprocessed spectrum helps constrain its
ionization, whereas in ASCA only the iron features can be used. The detailed
shape of the line and edge is a strong function of inclination as well as of
ionization. At high inclinations Doppler shifts prevail over gravitational and
transverse redshift, shifting the line and edge to higher energies as well as
giving substantial broadening. This is to zeroth order the same effect as
ionisation. However, the detailed shape of a low inclination, ionised
reflection spectrum is rather different to a less ionised reflection spectrum at
higher inclination, so these two parameters can be disentangled
given good enough data.

Figure 3a shows how $\chi^2$ varies as the inner disk inclination changes from
$30-66^\circ$ for the GSPC data, while Figure 3b shows this for the ASCA GIS 3
spectrum data. Clearly there is a significant preference in the data for an
inclination higher than $30^\circ$ (cosine smaller than 0.86), with a best fit
at $\sim 53^\circ$. This is rather higher than the inclination of $28-38^\circ$
inferred from optical studies, although within their firm upper limit of
$55^\circ$ (Sowers et al 1998). We fit all the EXOSAT and ASCA spectra with a
relativistic reprocessor inclined at $\sim 53^\circ$.  This gives similar
derived ionization for both sets of data, with typical values of $\xi\le 20$,
corresponding to $\le$ FeX. Figure 2b shows this high inclination
fit to the GIS 3 spectrum. This is a substantially better description
($\Delta\chi^2\sim 20$) of the data than the fit assuming a disk inclined at $30^\circ$. 

\begin{figure}[!h]
\begin{center} \leavevmode
\hbox{%
\epsfysize=0.45\textwidth
\epsffile[28 34 517 501]{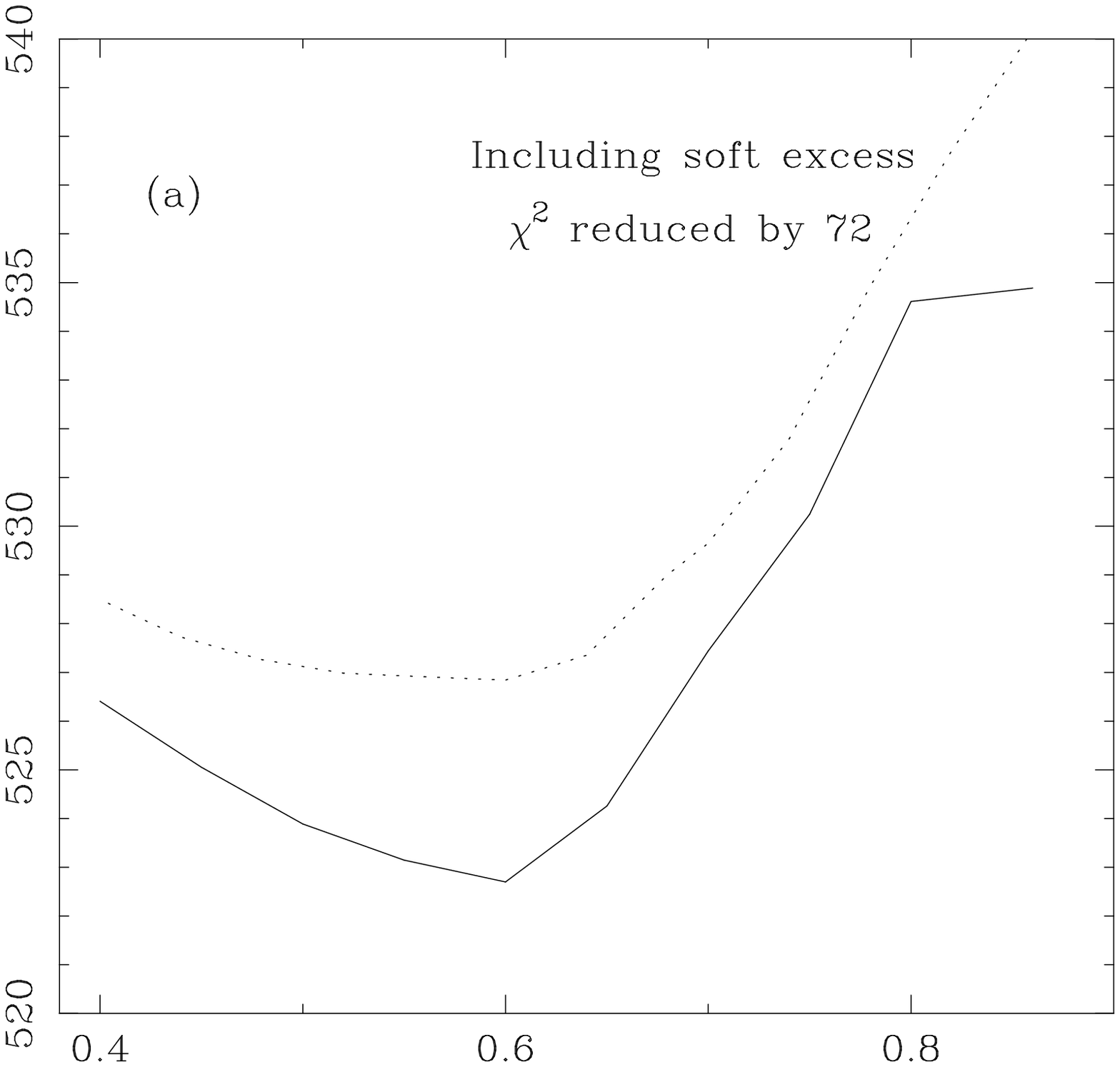}
\centering \leavevmode
\epsfysize=0.45\textwidth
\epsffile[28 34 517 501]{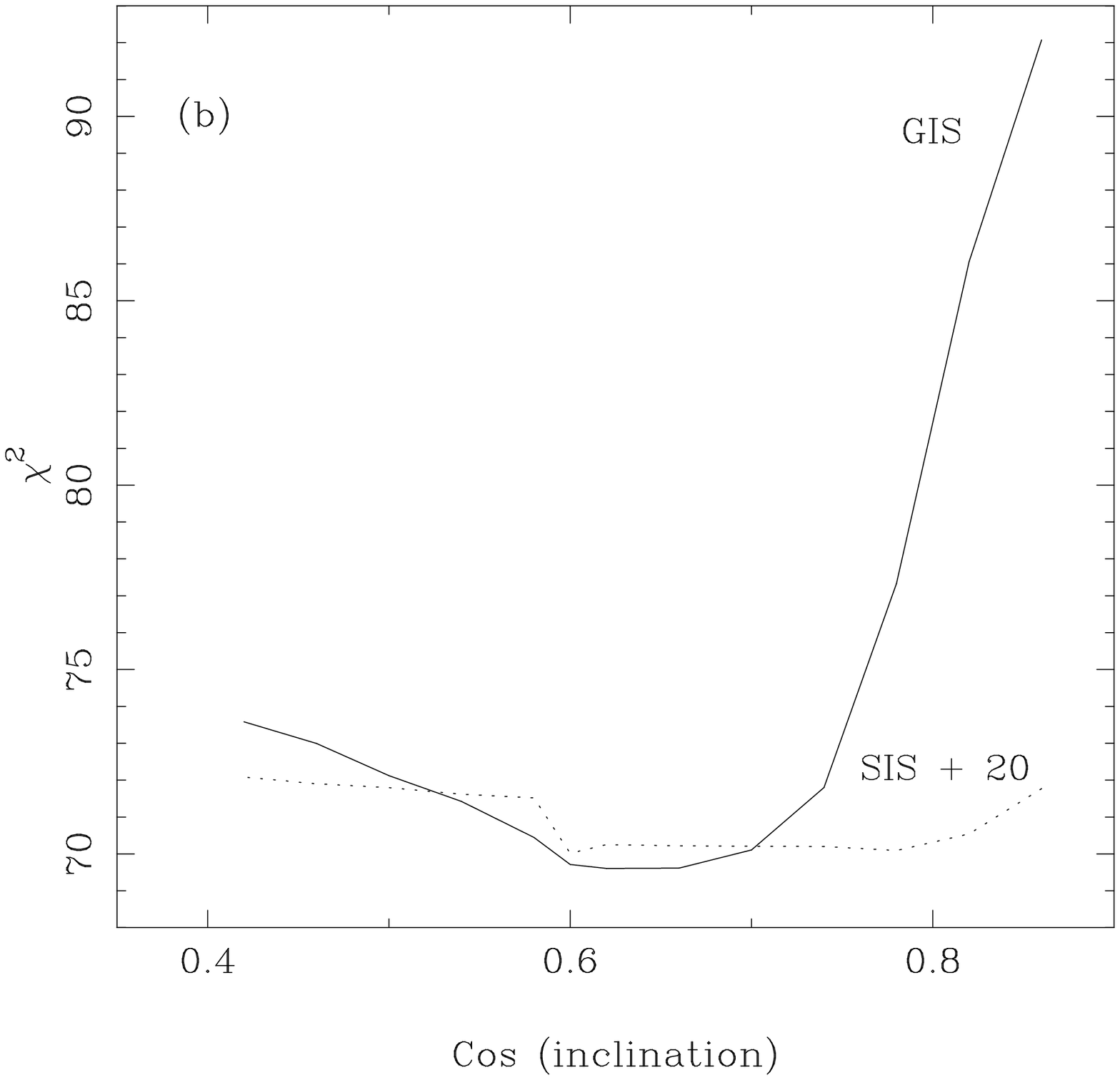}}
\end{center}
\caption{
The change in $\chi^2$ (goodness of fit parameter) as a function of inclination
of the reflecting material.  Panel (a) shows results from a simultaneous fit of
all the EXOSAT GSPC spectra, with the solid line showing fits to the data from
4--20 keV while the dotted line uses the full 2--20 keV bandpass including a
(diskblackbody) model for the soft excess. 
Panel (b) shows the ASCA GIS 3 (solid
line) and ASCA SIS 5 (dotted line) 4--10 keV spectral results.
}
\end{figure}

For both the EXOSAT and ASCA data, the high inclination fits have a derived
inner disk radius of $10-20R_g$, generally {\it inconsistent} with the
optically thick material extending down to the innermost stable orbit at
$6R_g$ (Done \& $\dot{\rm Z}$ycki 1998). A similar 'hole'
in the inner disk is also seen from the detailed reflected spectral shape 
of low/hard spectra from other GBHC ($\dot{\rm Z}$ycki et al., 1997; 1998ab),
implying that there is some robust physical mechanism which disrupts the flow.

\section{AGN: MCG--6--30--15}

We use our model to fit the seminal ASCA data from the AGN MCG--6--30--15, in
order to compare with our results from Cyg X--1 and the other GBHC.  Published
data usually concentrate on the ASCA SIS spectrum since this has the highest
spectral resolution. However, the time dependent corrections to the gain of the
SIS detectors due to radiation damage are not well understood and these problems
have been exacerbated by software errors
(http://heasarc.gsfc.nasa.gov/docs/asca/ rddrecipe.html).  Hence we fit the two
GIS spectra from MCG--6--30--15.
We use a spectral model with a power law continuum and its
relativistic reflection, together with two ionised absorbers in the line of
sight (see e.g. Otani et al 1997). Figure 4a shows the resulting fit which has
a rather steep intrinstic power law, with $\Gamma=2.11^{+0.14}_{-0.05}$,
an inner disk radius of $R_{\rm in}=6^{+1}R_{\rm g}$, and $\Omega/2\pi_{\rm
rel}=0.9^{+0.4}_{-0.3}$, with derived iron abundance of $3.7_{-2.3}^{+6.3}$
($\chi^2_\nu=1343/1308$). Plainly our code is capable of seeing a highly
relativistic disk, so the absence of the extreme relativistic components in Cyg
X--1 and the other GBHC is not due to our model assumptions.  As a cautionary
note, we also include the SIS 0 spectrum of MCG--6--30--15 compared to the best
fit GIS model (Figure 4b): the discrepancies are obvious.

\begin{figure}[!h]
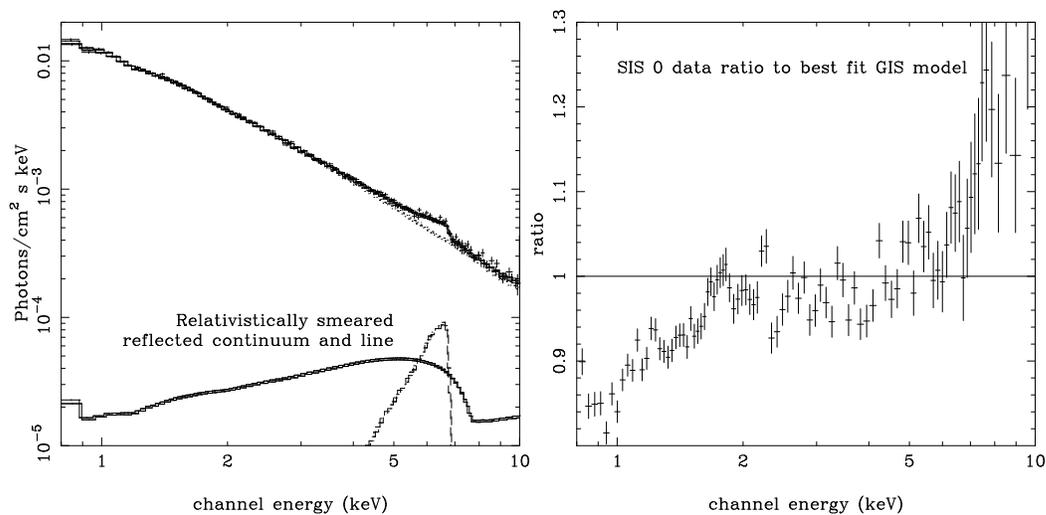

\begin{center} \leavevmode
\hbox{%
\epsfysize=0.45\textwidth
\epsffile[28 34 517 501]{done_4a.ps}
\centering \leavevmode
\epsfysize=0.45\textwidth
\epsffile[28 34 517 501]{done_4b.ps}}
\end{center}
\caption{
The left hand panel shows the best fit $30^\circ$ inclination relativistic disk
model for the GIS data from the AGN MCG--6--30-15.  The right hand panel shows
the ratio of this fit with the SIS 0 data.
}
\end{figure}

\section{Conclusions}

We significantly detect relativistic smearing of the reprocessed features in Cyg
X--1, and find that the disk is not highly ionised. The line is not suppressed
relative to the reflected continuum, it is merely broadened by the relativistic
effects which made it difficult to detect in ASCA data (E96), where the observed
weak and narrow line emission probably comes from the companion star.  The
derived covering fraction of the relativistic reflector is substantially less
than expected from a flat disk.  The most plausible explanation for these
observations is that the optically thin(ish) X--ray corona fills a central
`hole', so that less than half of the X--ray flux intercepts the disk. Similar
geometries are derived from energetic arguments about the continuum shape (see
e.g. Poutanen 1998). 

But what causes such a 'hole' in the inner disk ? The geometry is qualitatively
(but not quantitatively) consistent with the advective flow models proposed by
Esin et al., (1997; 1998). But is it really an advective flow ? 
The Esin et al (1997) models identify the soft--hard state transition with the
critical accretion rate at which the advective flows can exist. The disk then
has a constant efficiency at converting mass to radiation in the soft state
($L\propto\dot{M}$), which then changes to $L\propto\dot{M}^2$ for the advective
flow (as these have a lower radiative efficiency at lower density). The {\it
constant decay timescale} of the FRED soft X--ray transients seems then to
strongly argue against such models. And what happens to the inner disk in the
AGN ? Does it always extend down to $6R_g$, as in MCG--6--30--15 i.e. is the
inner disk instability supressed in AGN ? Or is MCG--6--30--15 an extremum in
the distribution of inner disk radii, as suggested by Zdziarski et al (1998).
What is the nature of the accretion flow onto a black hole ?

\section{References}

\re
   Basko M.M., 1978, ApJ., 223, 268
\re
   Done C., Mulchaey J.S., Mushotzky R.F., Arnaud K.A., 1992, ApJ.,395, 275
\re 
   Done C., \.{Z}ycki P. T.\ 1998, MNRAS, submitted
\re
   Ebisawa K., Ueda Y., Inoue H., Tanaka Y., White, N.E. 1996, ApJ., 467, 419
\re 
   Esin A. A., McClintock J. E., Narayan R.\ 1997, ApJ 489, 865
\re 
   Esin A. A., Narayan R., Cui W., Grove J. E., Zhang S.-N.\ 
1998, ApJ 505, 854
\re
   Fabian A.C., Rees M.J., Stella L., White, N.E. 1989, MNRAS, 238, 729
\re
   George I.M., Fabian A.C., 1991, MNRAS, 249, 352
\re 
   Gierli\'{n}ski M. et al.\ 1997, MNRAS 288, 958
\re 
   Kitamoto S., et al., 1984, PASJ, 36, 799
\re
   Lightman A.P., White T.R., 1988, ApJ, 335, 57
\re 
   Morrison R., McCammon D., ApJ., 270, 119
\re
   Otani C., et al., 1997, PASJ, 48, 211
\re
   Pounds K.A., et al., 1990, Nature, 344, 132
\re 
   Poutanen J.\ 1998, in Theory of Black Hole Accretion Discs,
   eds.\ M.\ A.\ Abramowicz, G.\ Bj\"{o}rnsson, J.\ E.\ Pringle 
  (CUP, Cambridge) (astro-ph/9805025)
\re
   Ross R.R., Fabian A.C., Brandt W.N., 1996, MNRAS, 278, 1082
\re 
   Sowers J.W., et al., 1998, ApJ., 505, 424
\re
   Tanaka Y. et al. 1995, Nature, 375, 659
\re
   Zdziarski A.A., Lubi\'nski P., Smith D.A., 1998, MNRAS, submitted
\re  
 \.{Z}ycki P. T., Done C.,  Smith D. A.\ 1997, ApJ 488, L113 
\re  
 \.{Z}ycki P. T., Done C.,  Smith D. A.\ 1998a, ApJ 496, L25 
\re  
 \.{Z}ycki P. T., Done C.,  Smith D. A.\ 1998b, MNRAS, submitted 
    (astro-ph/9811106)

\end{document}